\documentclass[JIP]{ipsj}
\usepackage{amsmath}
\usepackage{graphicx}
\usepackage{latexsym}
\usepackage{cite}
\usepackage{url}

\usepackage{framed}

\usepackage{color}
\definecolor{darkblue}{rgb}{0.0,0,0.5} 
\newcommand{\kw}[1]{{\tt {\color{darkblue} #1}}}
\newcommand{\nt}[1]{{\tt {\color{darkblue} #1}}}

\usepackage{preprint}
\PROheadtitle{To appear in Journal of Information Processing, 2015}

\def\Underline{\setbox0\hbox\bgroup\let\\\endUnderline}
\def\endUnderline{\vphantom{y}\egroup\smash{\underline{\box0}}\\}
\def\|{\verb|}

\received{2014}{04}{10}
\accepted{2015}{07}{30}

\usepackage[varg]{txfonts}
\makeatletter%
\input{ot1txtt.fd}
\makeatother%

\begin{document}

\title{Fast, Flexible, and Declarative Construction of Abstract Syntax Trees with PEGs}

\affiliate{YNU}{Graduate School of Electronic and Computer Engineering,
Yokohama National University,
79-1 Tokiwadai, Hodogaya-ku, Yokohama 240--8501 Japan}

\author{Kimio Kuramitsu}{YNU}[kimio@ynu.ac.jp]

\begin{abstract}
We address a declarative construction of abstract syntax trees with Parsing Expression Grammars. AST operators (constructor, connector, and tagging) are newly defined to specify flexible AST constructions. A new challenge coming with PEGs is the consistency management of ASTs in backtracking and packrat parsing. We make the transaction AST machine in order to perform AST operations in the context of the speculative parsing of PEGs. All the consistency control is automated by the analysis of AST operators. The proposed approach is implemented in the Nez parser, written in Java. The performance study shows that the transactional AST machine  requires 25\% approximately more time in CSV, XML, and C grammars.
\end{abstract}


\maketitle


\section{Introduction}

A parser generator is a standard method for implementing parsers in practical compilers and many other software engineering tools. The developers formalize a language specification with a declarative grammar such as LR($k$), LL($k$), GLR, or PEGs\cite{POPL04_PEG}, and then generate a parser from the formalized specification. However, in reality, many generated parsers are not solely derived from a formal specification. Rather, most parsers are generated with a combination of embedded code, called {\em semantic actions}.

The use of semantic actions has been a long tradition in many parser generators since the invention of yacc\cite{Yacc}. 
One particular reason is that a formal grammar itself is still insufficient for several necessary aspects of practical parser generation. The construction of {\em Abstract Syntax Trees} (ASTs)  is one of the such insufficient aspect of a formal grammar. Usually, the grammar developers write semantic actions to construct their intended form of ASTs. However, the semantic action approach lacks the declarative property of a formal grammar and reduces the reusability of grammars, especially across programming languages.

The purpose of this paper is to present a declarative extension of PEGs for the flexible construction of ASTs. The "declarative" extension stands for no semantic actions that are written in a general-purpose programming language. The reason we focus on PEGs is that they are closed under composition (notably, intersection and completion); this property offers better opportunities to reuse grammars.


We have designed AST operators that use an annotation style in parsing expressions, but allow for a flexible transformation of ASTs from a sequence of parsed strings. The structures that we can transform include a nested tree, a flattened list, and left/right-associative pairs. Due to a special left-folding operator, the grammar developers can construct a tree representation for binary operators that keep their associativity correct.

We have addressed how to implement AST operators in the context of PEG's speculation parsing. The transactional AST machine is an machine abstraction of AST operations, in which the intermediate state of ASTs while parsing is controlled at each fragment of mutation. The AST machine produces the resulting ASTs in either the full lazy evaluation way (such as in function programming) or the speculation way at any point of parsing expressions. Either ways, the produced AST is always consistent against backtracking. Synchronous memoization is presented as the integration of the AST machine with packrat parsing, in which the immutability of memoized results is ensured.

 
Recently, the use of parser generators has been extensively accepted for protocol parsers and text-data parsers\cite{ICDT11_PADS,OOPSLA14_ParserCombinator}. Parser performance, including the time cost of data extraction (i.e., AST construction in the parser terminology), is an integral factor in tool selection\cite{IMC06_Binpac,WWW06_XMLScreamer}. We have tested the Nez parser, which is implemented with the AST machine with the synchronous memoization. We demonstrate that the transactional AST machine approximately requires approximately 25\% more time in major grammars such as CSV, XML, and C. 

This paper proceeds as follows. 
Section 2 states the problem with AST constructions in PEGs. 
Section 3 presents our extended notations for AST construction. 
Section 4 presents the transactional AST machine that makes AST construction consistent with backtracking. 
Section 5 presents the integration of packrat parsing with the transactional AST machine. 
Section 6 presnets the performance study. 
Section 7 reviews related work. 
Section 8 concludes the paper. Our developed tools are open and available at \url{http://nez-peg.github.io/}.

\section{Problem Statement} 


\subsection{Semantic Actions}

PEGs, like other formal grammars, only provide syntactic recognition capability. This means that  the parsed result is just a Boolean value indicating whether an input is matched or not. To obtain detailed parsed results such as ASTs, the grammar developers need additional specifications to describe how to transform parsed results. 

{\em Semantic actions} are most commonly used in today's parser generators in order to program AST constructions with a fragment of embedded code in a grammar. Figure \ref{fig:action} shows an example of a semantic action written in Java, a host language of Rats\verb|!|\cite{PLDI06_Rats}. The embedded code \verb|{...}| is a semantic action,  combined with a generated parser at the parser generation time and invoked at the parsing time. 

\begin{figure}[htb]
\begin{framed}
{\small \begin{verbatim}
constant Action<Node> LogicalAndExpressionTail =
  "&&":Symbol right:BitwiseOrExpression {
    yyValue = new Action<Node>() {
      public Node run(Node left) {
        Node e = GNode.create("Expr",left, right);
        e.setLocation(location(yyStart));
        return e;
      }
    };
  }
\end{verbatim} }
\end{framed}
\caption{Example of AST constructions in Rats$!$ }
\label{fig:action}
\end{figure}

An obvious problem with semantic actions is that the grammar definition depends tightly on the host language of the generated parser. This results in a loss of opportunity for reuse in many potential parser applications such as IDEs and other software engineering tools since the developers often need to write another grammar from scratch.

\subsection{Consistency Problem}

The PEGs' flexibility come from the speculation parsing strategy. Typically, backtracking requires us to control the consistency management by means such as discarding some part of the constructed ASTs; otherwise, the ASTs may contain unnecessary subtrees that are constructed by backtracked expressions. In Figure \ref{fig:action}, for example, it is undecided whether a {\tt Node} object becomes a part of the final ASTs. The developer adds the {\tt Action} constructor for consistency when backtracking.  This problem is not new for PEGs but is common for semantic actions being executed in speculative parsers such as \cite{PPPJ03_Elkhound} and \cite{LDTA11_SemanticAction}. However, the consistency  still relies largely on the developer's management of semantic actions. 

Another consistency problem arises in packrat parsing\cite{ICFP02_PackratParsing}, a popular and standard technique for avoiding  PEGs' potential exponential time cost. Roughly, packrat parsing uses memoization for nonterminal calls, represented by $(N, P) \mapsto R$, where $N$ is a set of nonterminals in a grammar, $P$ is a parsing position over an input stream, and $R$ is a set of intermediate parsed results. As a part of the additional parsed results, we need to represent an intermediate state for ASTs, constructed at each nonterminal. More importantly, all memoized results have to be immutable in packrat parsing. Accordingly, we need to analysis the immutability of ASTs from the static property of grammars with semantic actions.

\subsection{Parsing Performance and Machine Abstraction}

Recently, the applications of formal grammars have been expanded from programming languages to protocol parsers and data analysis\cite{IMC06_Binpac,WWW06_XMLScreamer,OOPSLA14_ParserCombinator}. Parsing performance becomes a significant factor in  parser tool selection. In this light, semantic actions written in functional languages would provide a very consistent solution to the AST construction but not to our option.

One of the research goals of the Nez parser generator is high-performance parsing for ``Big Data'' analysis. 
In the context of text-data parsing, the AST construction roughly corresponds to data extraction and transformation tasks. 
For the sake of enabling dynamic grammar loading, the Nez parser generates not only parser source code but also byte-compiled code for the specialized parsing runtime. The machine abstraction is demanded for the AST construction instead of local variables and recursive calls in a recursive decent parsing.

\section{Extending AST Construction}

\subsection{ASTs}

An AST is a tree representation of the abstract structure of parse results. The tree is "{\em abstract}" in the  sense that it contains no unnecessary information such as white spaces and grouping parentheses. Figure \ref{fig:ast} shows an example of ASTs that are parsed from an if-condition-then expression. Each node has a {\em tag}, prefixed by \kw{\#}, to identify the meaning of the tagged node. A parsed substring is denoted by a single quotation \verb|' '|. For readability, we omit any parsed substrings in non-leaf nodes.  

For convenience, we introduce a textual notation of ASTs, which is exactly equivalent to the pictorial notation. Here is a textual version of Figure \ref{fig:ast}:

\begin{figure}[tb]
\includegraphics[width=8.0cm]{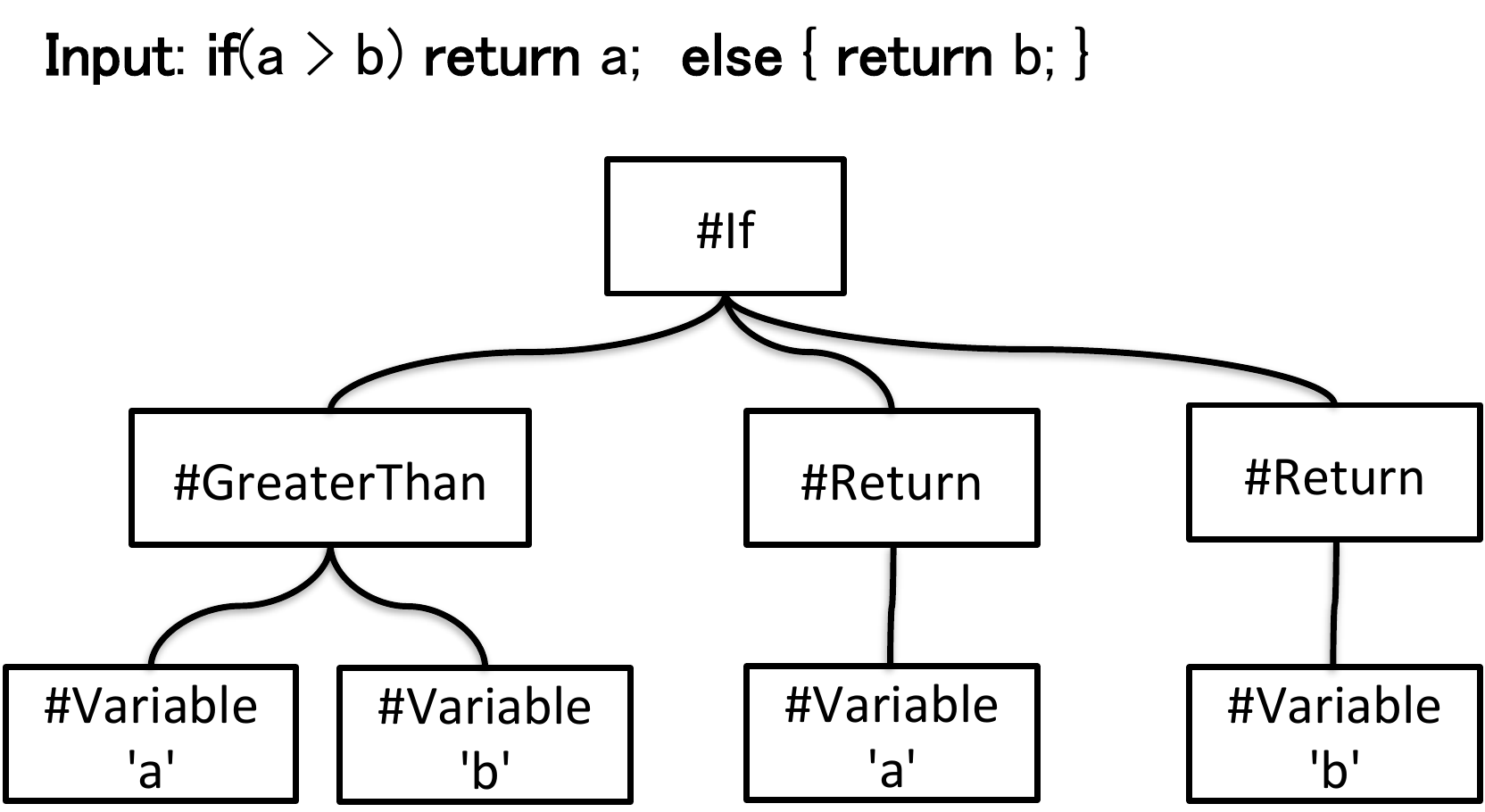}
\caption{Pictorial notation of ASTs}
\label{fig:ast}
\end{figure}

{\small \begin{verbatim}

#If[
  #GreaterThan[#Variable['a'] #Variable['b']]
  #Return[#Variable['a']]
  #Return[#Variable['b']]
] 
\end{verbatim}}

To be precise, the syntax of the textual notation of ASTs, denoted $T$, is defined inductively:

\[
 T ~ \verb|:==| ~~ \#t[T] ~~~ | ~~~ \#t[\verb|'...'|] ~~~ | ~~~T ~ T
\]

\noindent where $\#t$ is a tag to identify the meaning of $T$ and a parsed substring written by \verb|'...'|. A whitespace concatenates two or more nodes as a sequence. In this paper, we assume that the parsed result always starts with a non-sequence form of $\#t[T]$. 

Note that our AST definition is a minimalist; we drop any labeling for subnodes, like if(cond, then, else). While the labeling may be convenient when accessing subnodes, the sequence preserves the order of subnodes, providing sufficient semantics to distinguish them.  

\subsection{PEG Operators}

A PEG is a collection of productions, mapping from nonterminals to expressions. To write productions, we use the following form:
\[
A = e
\]
\noindent
where $A$ is the name of a nonterminal and $e$ is a parsing expression to be evaluated. Parsing expressions are composed by PEG operators. AST operators are designed to create and mutate ASTs in the parsing context of PEGs. Table \ref{table:peg4d} shows a summary of the PEG/AST operators. 

\begin{table}[bt]
\begin{center}
\caption{PEG/AST Operators} 
\label{table:peg4d}
\begin{tabular}{llccl} \hline
PEG  & Type & Operate & Proc. & Description\\ \hline
\verb|' '| & Primary & PEG & 5 & Matches text\\
\verb|[  ]| & Primary & PEG & 5& Matches character class \\
$.$ & Primary & PEG & 5 & Any character\\
$A$ & Primary & PEG & 5 & Non-terminal application\\
$\#t$ & Primary & AST & 5 & Tagging \\
$(~e~)$ & Primary & PEG & 5 & Grouping\\
$\{~e~\}$ & Primary & AST & 5 & Constructor\\
$\{@~e~\}$ & Primary & AST & 5 & Left-folding\\
$e?$ & Unary suffix & PEG & 4 & Option\\
$e*$ & Unary suffix & PEG & 4 & Zero-or-more repetitions\\
$e+$ & Unary suffix & PEG & 4 & One-or-more repetitions\\
$\&e$ & Unary prefix & PEG & 3 & And-predicate\\
$!e$ & Unary prefix & PEG& 3 & Negation\\
$@e$ & Unary prefix & AST& 3 & Connector \\
$e_1 e_2$ & Binary & PEG& 2 & Sequencing\\
$e_1 / e_2$ & Binary & PEG & 1 & Prioritized Choice\\ \hline
\end{tabular}
PEG: PEG operators, AST: AST operators
\end{center}
\end{table}

To begin, we recall the interpretation of PEG operators. The string \verb|'abc'| exactly matches the same input, while \verb|[abc]| matches one of these characters. The . operator matches any single character. The lexical match consumes the matched size of characters and moves forward a position of matching. The $e?$, $e*$, and $e+$ expressions behave as in common regular expressions, except that they are greedy and matches until the longest position. The $e_1 ~ e_2$ attempts two expressions $e_1$ and $e_2$ sequentially, backtracking the starting position if either expression fails.  The choice $e_1 / e_2$ first attempt $e_1$ and then attempt $e_2$ if $e_1$ fails. The expression $\&e$ attempts $e$ without any character consuming. The expression $!e$ fails if $e$ succeeds, but fails if $e$ succeeds. A more formal definition is detailed in \cite{POPL04_PEG}.

\subsection{AST Operators}

The design of the AST operators was inspired by the {\em substring capturing} commonly used in extended regular expressions such as Perl and PCRE\cite{Pcre}. Instead of \verb|( ... )|, we use $\{~e~\}$ to specify a substring that  we want to capture as an AST node. Here are two expressions that capture the same substring \verb|34| in an input \verb|123456|.

{\small \begin{verbatim}

  Regular Expression:  12(34)56
  Paring Expression: '12' { '34' } '56'

\end{verbatim}}

The major difference from the substring capturing in regular expressions is that we enhance the structural construction of nodes. To start, we introduce a global state reference, called the {\em left node}. The left node is implicit in notations but simply refers to an AST node that is constructed on the left hand of a parsing expression. To the left node, we define the following structural constructors:
 
\begin{itemize}
\item tagging, $\#t$ -- tagging the specified $\#t$ to the left node;
\item appending, $@e$ -- appending an $e$'s constructed node to the left node; and
\item connecting, $@[n]e$ -- setting an $e$'s constructed node at the $n$th position of child nodes on the left node
\end{itemize}

The tag $\#t$ is introduced to identify the meaning of nodes. Grammar developers are allowed to define a set of tags that they want. The tagging operator is used to specify such a tag on the left node. Untagged nodes are \verb|#tree| and \verb|#token| as default tags for tree nodes and leaf nodes respectively. 

\begin{figure}[tb]
{\small \begin{framed} \begin{verbatim}

  Value  = { [0-9]+ }
  Number = { [0-9]+ } #Int
  
\end{verbatim} \end{framed} }

{\small \begin{verbatim}

  Value    :: 12
        #token ['12']

  Number :: 12
        #Int ['12']
  
\end{verbatim}}
\caption{Example of tagging and its constructed AST nodes}
\label{fig:tagging}
\end{figure}

Figure \ref{fig:tagging} shows an example of tagging. We use $A :: s$ to represent an input $s$ for the production $A$. Since the left node is a global state, we can specify the tagging across nonterminals. In addition, the new left node is set at the position of opening brace \verb|{|. The {\tt Number} production can be equally specified in the following ways:

{\small \begin{verbatim}

  Number = Value #Int
  Number = { #Int [0-9]+ } 
  Number = { [0-9]+ #Int} 
  
\end{verbatim}}

An annotation style of tagging is introduced to be flexible for the meaning of nodes depending on the parse results. Consider the following case where the type of numbers are decided on the suffix \verb|[Ll]| followed by numbers. We can specify the structure of AST nodes  without any modification of the original parsing expressions \verb|[0-9]+ [Ll]?|. Note that duplicated tagging is regarded as overridden. 

{\small \begin{verbatim}

  Number = { [0-9]+ #Int ([Ll] #Long)? } 
  
\end{verbatim}}

The $@e$ operator connects two nodes in a parent-child relation. The prefix $@$ is used to specify a child node and append it to the left node as the parent. This is followed by the natural order of the top-down parsing. In addition, we allow the indexer, denoted $@[n]$, in order to specify the $n$th position of the child node on the left node. Figure \ref{fig:tree} is an example of tree transformation with/without a repetition. As shown in {\tt AdditiveM} the {\tt @Number} inside the repetition is regarded as a natural addition of nodes, while the {\tt @[1]Number} inside the repetition overrides the second nodes over again.  

\begin{figure}[tb]

{\small \begin{framed} \begin{verbatim}

  Additive = { @Number '+' @Number #Add }
  Additive2 = { @[1]Number '+' @[0]Number #Add }

  AdditiveM = { @Number ('+' @Number)+ #Add }
  AdditiveM2 = { @Number ('+' @[1]Number)+ #Add }
  
\end{verbatim}\end{framed}}

{\small \begin{verbatim}

  Additive  :: 1+2
        #Add[ #Int['1']  #Int[2] ]

  Additive2 :: 1+2
        #Add[ #Int['2']  #Int[1] ]

  AdditiveM ::  1+2+3+4
        #Add[ #Int['1']  #Int[2] #Int[3] #Int[4] ]

  AdditiveM2 ::  1+2+3+4
        #Add[ #Int['1']  #Int[4]  ]
  
\end{verbatim}}

\caption{Example of tree constructions by @e}
\label{fig:tree}
\end{figure}

Note that the $@e$ operator works under the assumption that a new node is created by $e$. In reality, many grammar developers often connect an uncreated expression. In this case, we treat it as an error due to avoiding a cyclic structure by self-referencing nodes. More importantly, the error can be easily detected at runtime by comparing the left node of $@e$ and the result node of $e$. If both nodes are the same, we ignore such an erroneous connection. 

\subsection{Left Folding}

In the previous subsection, we present a tree construction with AST operators. Basically, the AST operators can transform the parsed substring into a tree structure. That is, we can specify whether a subtree is either nested, flattened, or ignored. As shown in Figure \ref{fig:list} and \ref{fig:right}, we make the construction of a flattened list and right-associative pairs from a sequence A,B,C,D. On the other hand, we have to pay a special attention to the construction of left-associative pairs. For example, the following is the construction of a left-associative paris although the grammar contains left-recursion.

{\small \begin{verbatim}

 Expr = Pair / Term
 Pair =  {@Expr ',' @Term #Pair }   // left-recursion!!
 Term = { [A-z]+ #Term }

\end{verbatim}}

Left-recursion is a major restriction of PEG. Although there is a known algorithm for eliminating any left-recursion from a grammar (as shown in \cite{OOPSLA14_Antlr}), this elimination does not ensure the left associativity. 

\begin{figure}[tb]
{\small \begin{framed} \begin{verbatim}
Expr = List / Term
List  = { @Term (',' @Term)+ #List}
Term = {[A-z] #Term}
\end{verbatim} \end{framed}}
\caption{Construction of a flattened list [A, B, .., C, D]}
\label{fig:list}
\end{figure}

\begin{figure}[tb]
{\small \begin{framed} \begin{verbatim}
Expr = Pair / Term
Pair =  {@Term ',' @Expr #Pair } 
Term = { [A-z] #Term }
\end{verbatim} \end{framed} }
\caption{Construction of right-associative pairs [A, [B, ... [[C,D]]]]}
\label{fig:right}
\end{figure}

\begin{figure}[tb]
{\small \begin{framed} \begin{verbatim}
Expr = Term {@ (',' @Term) #Pair }*
Term = {[A-z] #Term}
\end{verbatim} \end{framed} }
\caption{Construction of left-associative pairs [[[A, B], ..., C], D] with left-folding}
\label{fig:left}
\end{figure}

{\em Left-folding} is additionally defined as constructing a left-associative structure from the repetition.  Left-folding $\{@ ~ e\}$ is creating a new node that contains the left node as the first child node. That is, $e_1 \{@ ~ e_2\}$ is equivalent to $\{@e_1 ~ e_2\}$. Usually, we use the left-folding with a repetition ($e_1 \{@ ~ e_2\}*$), or  $(e_1 ... \{@ ~ e_2\} )  \{@ ~ e_2\} $. Note that $e_1 \{@ ~ e_2\}*$ is equivalent to  $A = \{@A ~ e_2\} ~ / ~ e_1$ although $A$ is left-recursive. 

Figure \ref{fig:left} is the construction of left-associative pairs from A,B, ..., C, D. As the name implies, left-folding is chiefly used for constructing left-associative binary operators. Figure \ref{fig:math} shows the basic mathematic operations with AST operators. 

\begin{figure}[tb]
\begin{tabular}{ll} \hline
{\tt Expr} & \verb|= Sum| \\
{\tt Sum} & \verb|= Product {@ ( '+' #add / '-' #sub ) @Product }*| \\
{\tt Product} & \verb|= Value {@ ( '*' #mul / '/' #div) @Value }*| \\
{\tt Value} & \verb|= { [0-9]+ #Integer } / '(' Expr ')' | \\ \hline
\end{tabular}
\caption{Example of Basic Mathematical Operators }
\label{fig:math}
\end{figure}


\subsection{Operational Semantics}


\begin{figure}[tb]
\begin{center}
\begin{tabular}{|lrll|} \hline
$e$ &  \verb|      ::= | & $\epsilon$ & : empty \\ 
& \verb#|  # & $A$ & : nonterminal  \\
& \verb#|  # & \verb|'a'| & : terminal character \\
& \verb#|  # & $e$ $e' $ &  : sequence \\
& \verb#|  # & $e {\tt /} e' $ &  : prioritized choice \\
& \verb#|  # & $e?$ &  : option $e/\epsilon$  \\
& \verb#|  # & $e*$ &  : repetition $A = e A /\epsilon$ \\
& \verb#|  # & \verb|&|$e$ & : and predicate  \\
& \verb#|  # & \verb|!|$e$ & : not predicate  \\
& \verb#|  # & $\{ e \}$ & : constructor  \\
& \verb#|  # & $@e$ & : linking child  \\
& \verb#|  # & $\{@ ~e \}$ & : left-folding \\
& \verb#|  # & \verb|#T| & : tagging  \\
\hline
\end{tabular}
\end{center}
\caption{Syntax of PEGs with AST operators}
\label{fig:syntax}
\end{figure}

Finally, we define the operational semantics of AST operators in parsing expressions. To begin, we define several notations used in the semantics. Let $x,y,z \in \Sigma^*$ be a sequence of characters and $xy$ be a concatenation of $x$ and $y$. We write $T$ for a node of ASTs. $\#{\tt t}[x]$ is a newly created node with a default tag $\#{\tt t}$ and a substring $x$. $T[T']$ stands for adding a child $T'$ to the parent $T$. $T/\#t$ stands for the replacement of the tag of $T$ with the specified $\#t$.

The semantics of $e$ is defined by a state transition $(xy, T)  \xrightarrow{e} (y, T')$, which can be read: the expression $e$ parsing the input stream $xy$ consumes $x$ and transforms the left node $T$ into $T'$. If $T = T'$ in the transition, then the node is not mutated.

Figure \ref{fig:syntax} is an abstract syntax of parsing expressions with the AST operators. Due to space constraints, we highlight core parsing expressions, which only contain $\epsilon$, $a$, $e_1 ~ e_2$, $e_1 / e_2$, and $!e$. Other expressions, including character class, option, repetition, and-predicate, can be rewritten by these core expressions \cite{POPL04_PEG}. Without the loss of generality, we omit $@[n]e$ from the syntax definition. Figure \ref{fig:sem} shows the definition of the operational semantics of $e$. We write $\bullet$ for a special failure state. Any transitions to $\bullet$ suggests the backtracking to the alternative if one exists. 

\begin{figure}[tb]

{\small 
\[
     (x, T)  \xrightarrow{\epsilon} (x, T) 
\]

\[
\frac{a}
     { (ax, T)  \xrightarrow{a} (x, T)  } 
, ~~~
\frac{a ~~~ a \ne b}
     { (bx, T)  \xrightarrow{a} \bullet  } 
\]

\[
\frac{(xyz, T) \xrightarrow{e_1} (yz, T') ~~~ (yz, T')  \xrightarrow{e_2} (z, T'') }
     { (xyz, T)  \xrightarrow{e_1 e_2} (z, T'')  }
, ~~~
\frac{(xyz, T) \xrightarrow{e_1} \bullet }
     { (xyz, T)  \xrightarrow{e_1 e_2} \bullet  } 
\]

\[
\frac{ (xyz, T) \xrightarrow{e_1} (yz, T') ~~~ (yz, T')  \xrightarrow{e_2} \bullet }
     { (xyz, T)  \xrightarrow{e_1 e_2} \bullet  } 
\]

\[
\frac{(xy, T) \xrightarrow{e_1} (y, T')  }
     { (xy, T)  \xrightarrow{e_1 / e_2} (y, T')  } 
, ~~~
\frac{(xy, T) \xrightarrow{e_1} \bullet ~~~ (xy, T)  \xrightarrow{e_2} (y, T') }
     { (xy, T)  \xrightarrow{e_1 / e_2} (y, T')  } 
\]

\[
\frac{(xy, T) \xrightarrow{e_1} \bullet ~~~ (xy, T)  \xrightarrow{e_2} \bullet }
     { (xy, T)  \xrightarrow{e_1 / e_2} \bullet  } 
\]

\[
\frac{(x, T) \xrightarrow{e} \bullet }
     { (x, T)  \xrightarrow{!e} (x, T)  } 
, ~~~
\frac{(x, T) \xrightarrow{e} (x, T) }
     { (x, T)  \xrightarrow{!e} \bullet  }      
\]

\[
\frac{(xy, T) \xrightarrow{e} (y, T) }
     { (xy, T)  \xrightarrow{\{ e \}} (y, \#{\tt t}[x])  } 
, ~~~
\frac{(p, x) \xrightarrow{e} \bullet }
     { (p, x)  \xrightarrow{\{ e \}} \bullet  }      
\]

\[
      (x, T)  \xrightarrow{\#t} (x, T/\#t)
\]

\[
\frac{(xy, T) \xrightarrow{e} (y, T') ~~~ T \ne T' }
     { (xy, T)  \xrightarrow{@e} (y, T[T'])  } 
, ~~~
\frac{(xy, T) \xrightarrow{e} (y, T') ~~~ T = T' }
     { (xy, T)  \xrightarrow{@e} (y, T)  } 
, ~~~
\frac{(xy, T) \xrightarrow{e} \bullet }
     { (xy, T)  \xrightarrow{@e} \bullet  }      
\]

\[
\frac{(xy, T) \xrightarrow{e} (y, T') }
     { (xy, T)  \xrightarrow{\{@ ~ e\}} (y, \#{\tt t}[x][T])  } 
, ~~~
\frac{(xy, T) \xrightarrow{e} \bullet }
     { (xy, T)  \xrightarrow{\{@ ~ e\}} \bullet  }      
\]

}
\caption{Operational Semantics}
\label{fig:sem}
\end{figure}

PEG operators and AST operators are orthogonal to each other. In other words, AST operators do not influence the operational semantics of PEG operators. On the contrary, AST operators only use a substring that is matched by an expression $e$.

\section{Transactional AST Machine}

A transactional AST machine is a machine-based implementation to make the  AST construction consistent with AST operators. All operations are recorded as instruction logs to be canceled when backtracking. In this section, we describe the transactional AST machine.

\subsection{Machine and Instructions}

For simplicity, we start by assuming the absence of backtracking. An AST machine has the following three states: 

\begin{itemize}
\item $p$, a parsing position at which the parser attempts next,
\item $\textrm{left}$, a node reference to the {\em left node} that is operated by AST operators, and
\item a {\em node stack} to store a parent child relation of AST nodes
\end{itemize}

The AST machine provides the following instructions to operate the above three states:

\begin{itemize}
\item $\kw{push}(\textrm{left})$ -- push the $left$ node onto the node stack
\item $\textrm{left} \leftarrow \kw{pop}$ -- pop the top node as $\textrm{left}$
\item $\textrm{left} \leftarrow \kw{new}$ -- create an new node as $\textrm{left}$
\item $\kw{open}(\textrm{left},p)$ -- set $p$ as the starting position of $\textrm{left}$
\item $\kw{close}(\textrm{left},p)$ -- set $p$ as the ending position of $\textrm{left}$
\item $\kw{tag}(\textrm{left},s)$ -- tag the $\textrm{left}$ node with the specified $s$ 
\item $\kw{link}(\textrm{left})$ -- link the $\textrm{left}$ node into the stack top node
\item $\textrm{left} \leftarrow \kw{swap}(\textrm{left})$ -- swap the $\textrm{left}$ node and the stack top node
\end{itemize}

Note that the substring of a node is represented with the starting position and the end position over the input stream. 

The PEG parser moves over an input stream. From viewpoint of the AST machine, the parser itself can be viewed as a {\em blackbox} function. We write $p \leftarrow \kw{parse}(e, p)$ for the parser function --- parsing the input stream with $e$ where the character consumption is represented by its resulting moved position of $p$.

Let $\tau(e)$ be a compile function that converts from parsing expressions to a sequence of AST instructions. The function $\tau(e)$ is defined inductively in Figure \ref{fig:tau}.

\begin{figure}[tb]

\begin{tabular}{lrl} 
$\tau(\{~e~\})$ &  \verb|      = | & $\textrm{left} \leftarrow \kw{new}$  \\ 
 &  & $\kw{open}(\textrm{left}, p)$  \\
 &  & $\tau(e)$  \\
 &  & $\kw{close}(\textrm{left}, p)$  \\
$\tau(\#t)$ &  \verb|      = | & $\kw{tag}(\textrm{left}, {\tt t})$  \\ 
$\tau(@e)$ &  \verb|      = | & $\kw{push}(\textrm{left})$  \\ 
 &  & $\tau(e)$  \\
 &  & $\kw{link}(\textrm{left})$  \\
 &  & $\textrm{left} \leftarrow \kw{pop}$  \\
$\tau(\{@ ~e ~\})$ &  \verb|      = | & $\kw{push}(\textrm{left})$  \\ 
 &  & $\textrm{left} \leftarrow \kw{new}$  \\ 
 &  & $\textrm{left} \leftarrow \kw{swap}(\textrm{left})$  \\ 
 &  & $\kw{link}(\textrm{left})$  \\ 
 &  & $\kw{open}(\textrm{left}, p)$  \\
 &  & $\tau(e)$  \\
 &  & $\kw{close}(\textrm{left}, p)$  \\
$\tau(e)$ &  \verb|      = | & $p \leftarrow \kw{parse}(e, p)$  \\ 
\end{tabular} 
\caption{Definition of a compile function for AST machine}
\label{fig:tau}

\end{figure}

The compiled instructions ensure that the stack top is always a parent node at the execution time of \kw{link}. This is easily confirmed in the way that the \kw{link} instruction is compiled between \kw{push}(\textrm{left}) and \kw{pop}.

\subsection{AST Construction with Backtracking} \label{sec:sync}

Backtracking requires the rollback handling of the instruction executions, since some executions could be unnecessary when backtracking. Suppose $\{ ~ \#t ~ e_1 ~ \} ~/ ~e_2$, for example. Before evaluating $e_1$, we need three AST instructions (\kw{new}, \kw{open}, and \kw{tag}) to be executed. However, if the expression $e_1$ fails, these instructions are unnecessary before attempting alternatives $e_2$.

The transactional AST machine provides the {\em lazy} evaluation mechanism for the execution of AST instructions. The lazy evaluation means that we cannot perform any instructions until we reach a point where backtracking no longer occurs.  

The lazy evaluation can be simply achieved by logging instructions in a {\em stack-based buffer}. Let $i$ be a position of the latest stored instruction log on the buffer. The buffer is operated by the following transactional instructions:

\begin{itemize}
\item \kw{{\it log}} $\kw{push}|\kw{pop}|..|\kw{swap}$ -- log an AST instruction to the instruction buffer ($i = i + 1$);
\item $t \leftarrow \kw{save}$ -- save $i$ for the beginning of a transaction ($t = i$);
\item $\kw{commit}(t)$ -- execute instruction logs stored between $t$ and $i$ ($i = t$), and then expire them; and 
\item $\kw{abort}(t)$ -- expire the instruction logs stored between $t$ and $i$ ($i = t$).
\end{itemize}

In the above, we take an instruction form to represent the transactional operations. This is based on the implementation of a Nez interpreter-based parser. In practice, one could not necessarily implement these operations as instructions. Instead, the AST machine only provides APIs to control the \kw{save}, \kw{commit} and \kw{abort} operations for the parser. 

The \kw{abort} operation is fully automated on a PEG parser.  At the time of any failure occurrences, the parser aborts the transaction to the save point $t$. The save points are exactly the same points where the parser saves a parser position (over the input) to attempt alternatives when backtracking. To be precise, the $\Downarrow$ below indicating the save point for the transaction.

\begin{itemize}
\item $(\Downarrow e) / e'$,
\item $(\Downarrow e)$?,
\item $(\Downarrow e)$*,
\item $\& ( \Downarrow e )$,
\item $! ( \Downarrow e )$
\end{itemize}

Now we may commit the transaction at the any point in parsing expressions. However indiscriminate commitments may result in the speculative AST instantiation if backtracking occurs. As discussed in the next section, the speculative instantiation is also consistent against backtracking, although no unused instantiation is ideal. It is still unknown whether a certain point of the parser context never backtrack. The simplest solution is to invoke the \kw{commit} when the whole input is parsed. This gives us the benefit of a full lazy evaluation as in functional programming languages.  

\section{Packrat Parsing with ASTs}

Packrat parsing\cite{ICFP02_PackratParsing} is an essential technique to avoid the potential exponential time cost of backtracking. This section describes the safe integration of  the transactional AST machine with packrat parsing.

\subsection{Laziness vs. Speculation}

Packrat parsing \cite{ICFP02_PackratParsing} is a memoization version of the recursive decent parsing. Since all the intermediate parse results of nonterminal calls are memoized at each distinct position, we can avoid redundant calls, which lead to exponential time costs in the worst case. In the context of AST constructions, we additionally need to memoize the intermediate state of ASTs.

We consider two strategies: {\em lazy-full} and {\em speculation}. The lazy-full strategy involves memoizing instruction logs to take full advantage of lazy evaluation. The speculation strategy involves memoizing an AST node that is instantiated despite the fact that the instantiated node may eventually be unused and discarded. We choose the speculation strategy, after the following comparison of the pros and cons of both strategies. 

The lazy-full strategy is natural and very compatible with the transactional AST machine. An obvious advantage is that we take full advantage of lazy evaluation of AST constructions. However, a disadvantage is also clear;  we need to copy a large number of instruction logs to be memoized. Although the memoized logs can be reduced to a subsequence of logs that are only added by a given nonterminal, the size of the copy is roughly proportional to the size of input characters that the nonterminal has consumed. Since packrat parsing is based on the constant memoization cost in the size of the input, the memoized logs may invalidate the linear time guarantee. 

The advantage of the speculation strategy is that the reduced overhead of the memoization. Note that the instantiation costs of ASTs are not an actual overhead since we need the instantiation at least once even in the lazy-full strategy. Due to memoization, we can avoid the repeated instantiation of the same nodes. As a result, the overheads are the unnecessary instantiation and discard costs for the sake of eventually unused nodes. However, we consider that a modern garbage collector is efficient enough to handle such memory iterations. 

Another disadvantage is that we require the immutability analysis for the memoization point. To illustrate, we suppose that the production \nt{Symbol} that overrides the tag of a \nt{Name}-produced node.

{\small \begin{verbatim}

  Name = { NAME #Name }
  Symbol = Name #Symbol
  
\end{verbatim}}

Following the speculation strategy, an AST node is instantiated after \nt{Name} to be memoized. The same node can be memoized at \nt{Symbol}, but it is mutated by a different tagging \#Symbol. As a result, the lookup of the memoization table for \nt{Name} is different, as we have memoized at at \nt{Name}. 

In general, it is not easy to analyze the mutable region of nodes in parsing expressions with semantic actions. Fortunately, AST operators have restricted semantics in terms of the mutation of nodes. In addition, there is no method to mutate to a child node of the left node. Accordingly, the mutable region is surrounding by $@(e)$.

\subsection{Synchronous Memoization}

The memoization of an AST node is performed not at arbitrary nonterminals, but at a safe point where we ensure that the instantiated node is immutable. Let $m_i$ be an identifier that uniquely represents such a memoization point. Let $s_i$ be a starting point for the instantiation of the node for $m_i$. 

Synchronous memoization is a memoization that synchronizes with a transactional instantiation of an AST node. The following pseudo code illustrates the algorithm of the synchronous memoization of $(s_i, m_i)$. 

  $ .. $\\
  $left = Lookup(m_i)$ \\
  $if(\textrm{l is not found}) \{ $\\
  $\verb|    | /* s_i : \textrm{begin of transaction}*/$\\
  $\verb|    |\textrm{left is created and the mutated}$\\
  $\verb|    | /* m_i : \textrm{end of transaction} */$\\
  $\verb|    |left = Commit(s_i, m_i)$\\
  $\verb|    |Memoize(m_i, left);$ \\
  $\}$\\
  
Before the instantiation of a node, we use $Lookup(m_i)$ to find an already instantiated node from the memoization table. If found, we set it to the left node and never attempt any mutations for the set node. Otherwise, we start a transaction that instantiates a new node. During the transaction, the node mutations are all logged in the transactional AST machine. When backtracking occurs, the mutations are automatically aborted. If we reach at the $m_i$ point, we commit the logged instructions by $Commit(s_i, m_i)$ and then obtain an instantiated node. $Memoize(m_i)$ is called to store the instantiated node in the memoization table. 

\begin{figure}[tb]
\begin{tabular}{p{1cm}rrl} 
$\tau(@[n]e)$      &  \verb|      = | & & $\kw{push}(\textrm{left})$  \\
 &  & & $\textrm{left} \leftarrow \kw{lookup}(m)$  \\
 &  & & $\kw{ifnon}(\textrm{left}, L)$  \\
 &  & & $t \leftarrow \kw{save}()$  \\
 & & & $\tau(e)$  \\
 &  & & $\textrm{left} \leftarrow \kw{commit}(t)$  \\
 &  & & $\kw{memo}(m, \textrm{left})$  \\
 &  & L & $\kw{link}(\textrm{left})$  \\
 &  &  & $\kw{pop}(\textrm{left})$  \\ 
\end{tabular}
\caption{Synchronous memoization version of $\tau(@e)$}
\label{fig:memolink}
\end{figure}

Figure \ref{fig:memolink} illustrates the synchronous memoization version of $\tau(@e)$. The memoization point $m$ is an unique number for every distinct subexpression $e$, which is derived from the grammar analysis. 

Note that nonterminal calls in general are not memoized in the synchronous memoization. However, this may reduce the number of memoization points and decrease the effect of packrat parsing. On the other hand, nonterminals involving no AST operations have no side effect for node constructions. In the Nez parser, we use such nonterminals for another available memoization point. 

\subsection{Garbage Collection}

Another problem with the speculation strategy is how to discard unused nodes. 
Unused nodes inevitably occur since the instantiated nodes are temporarily stored on the {\tt link} logs before their parent nodes are instantiated. (Note that the {\tt link} logs can be always expired by backtracking). 
The memoization table on the other hand has to keep the expired nodes from the logs in order to avoid the reinstantiation of the same node.

The conventional packrat parsers keep all memoized results until the whole parser process ends\cite{ICFP02_PackratParsing}. This suggests that the heap consumption considerably increases when we add all intermediate AST nodes. Worse, it is impossible in general to determine the point at which a memoized node is no longer used\cite{PASTE10_Yapp}. 

One practical solution is the use of a sliding window to range the memoization table over the input position. In the sliding window, memoized nodes are expired if the parse moves forward in the window size. Our previous work \cite{PRO101} confirms both the linear time parsing and the constant memory consumption if the window size is large enough to cover the length of backtracking. The Nez parser uses the sliding window for memoization, and allows the garbage collector to collect expired nodes. This results in the reduced memory pressure.  

\begin{table*}[tb]
\caption{Summary of Grammars and Data Sets}
\label{table:grammars}
\begin{tabular}{llrr|rrrrrrr} \hline
Grammars & Operation &Productions & Memo Points & File Size & Latency [ms] & Backtrack &  Memo Effects & \# of Nodes & \# of Unused \\ \hline
CSV & PEG & 3 & 2 & 8.8MB & 914 & 0 &  0 &  &  \\
CSV & AST &  3 & 1 & 8.8MB & 2102 & 0 &  0 & 1,480,777 & 0 \\
XML & PEG & 14 & 7 & 11.6MB & 1276 & 0.03528 & 0 &  &  \\
XML & AST  & 14 & 5 & 11.6MB & 1777 & 0.03528 &  0 & 568,668 & 0 \\
C & PEG & 153 & 120 & 1.31MB & 857 & 0.96852 &  0.13472 &  &  \\
C & AST & 153 & 110 & 1.31MB & 1193 & 0.97237 &  0.38985 & 160,301 & 66,366 \\
JS & PEG & 127 & 49 & 247KB & 294 & 12.689 &  0.26216 &  &  \\
JS & AST  & 127 & 58 & 247KB & 499 & 15.43039 &  0.69062 & 32,475 & 97,993 \\ 
\hline
\end{tabular}
\end{table*}  

\section{Experimental Results}

This section describes the results of our performance study on AST constructions on the Nez parser. 

\subsection{Parser Implementation}

Nez is a PEG-based parser generator that has a language support for the AST operators.
The Nez parser is written in Java, and integrated with enhanced packrat parsing with sliding window, presented in \cite{PRO101}, and the transactional AST machine with synchronous memoization, described in Sections 4 and 5. 

In this experiment, we run the Nez parser as an interpreter mode, although it can generate parser source code. The Nez interpreter is highly optimized with several techniques including grammar inlining, partial DFA-conversions, and superinstructions. 

The test environment is Apple Mac Book Air, with 2GHz Intel Core i7, 4MB of L3 Cache, 8GB of DDR3 RAM, on Mac OS X 10.8.5 and Oracle Java Development Kit version 1.8. All measurements represent the best result of five or more iterations over the same input.

\subsection{Grammars and Datasets}

The grammars we have investigated are selected from the same set \cite{PRO101} in such a way that we can examine the variety of backtracking activity. Data sets are chosen to demonstrate a typical parser behavior for the given grammar. We label the pair of tested grammar and dataset as follows.

\begin{itemize}
\item CSV -- a simple grammar that involves no backtracking and many flattened AST nodes. The tested data come from an open data file offered by the JapanPost.  
\item XML -- a typical grammar for data formats that involves low backtracking activity and many nested AST nodes. The tested data are obtained from the XMark benchmark project\cite{VLDB02_XMark}. 
\item C -- a language grammar that involves moderate backtracking activity. The tested data are derived from Google NSS Cache project. 
\item JS -- a language grammar that involves high backtracking activity and then shows an exponential time cost, as reported in \cite{PRO101}. The tested data are an uncompressed jquery source file.  
\end{itemize}

Table \ref{table:grammars} shows a summary of grammars and datasets. The left side of the table indicates the static properties of grammars. The column labeled "Production" stands for the number of productions, and Column "Memo Points" stands for the number of memo points. The right side of the table indicates the statistics of internal parser behaviors when we parse the data sets. Column "Backtrack" stands for the backtrack activity, measured by the ratio of the total backtracking length by the input size. Column "Memo Effects" is measured by the hit ratio of memoized results. Column "Nodes" stands for the number of nodes that the final ASTs contain, and Column "Unused" stands for the number of eventually unused nodes.

\subsection{Performance Study}

\begin{figure}[tb]
\includegraphics[width=8.0cm]{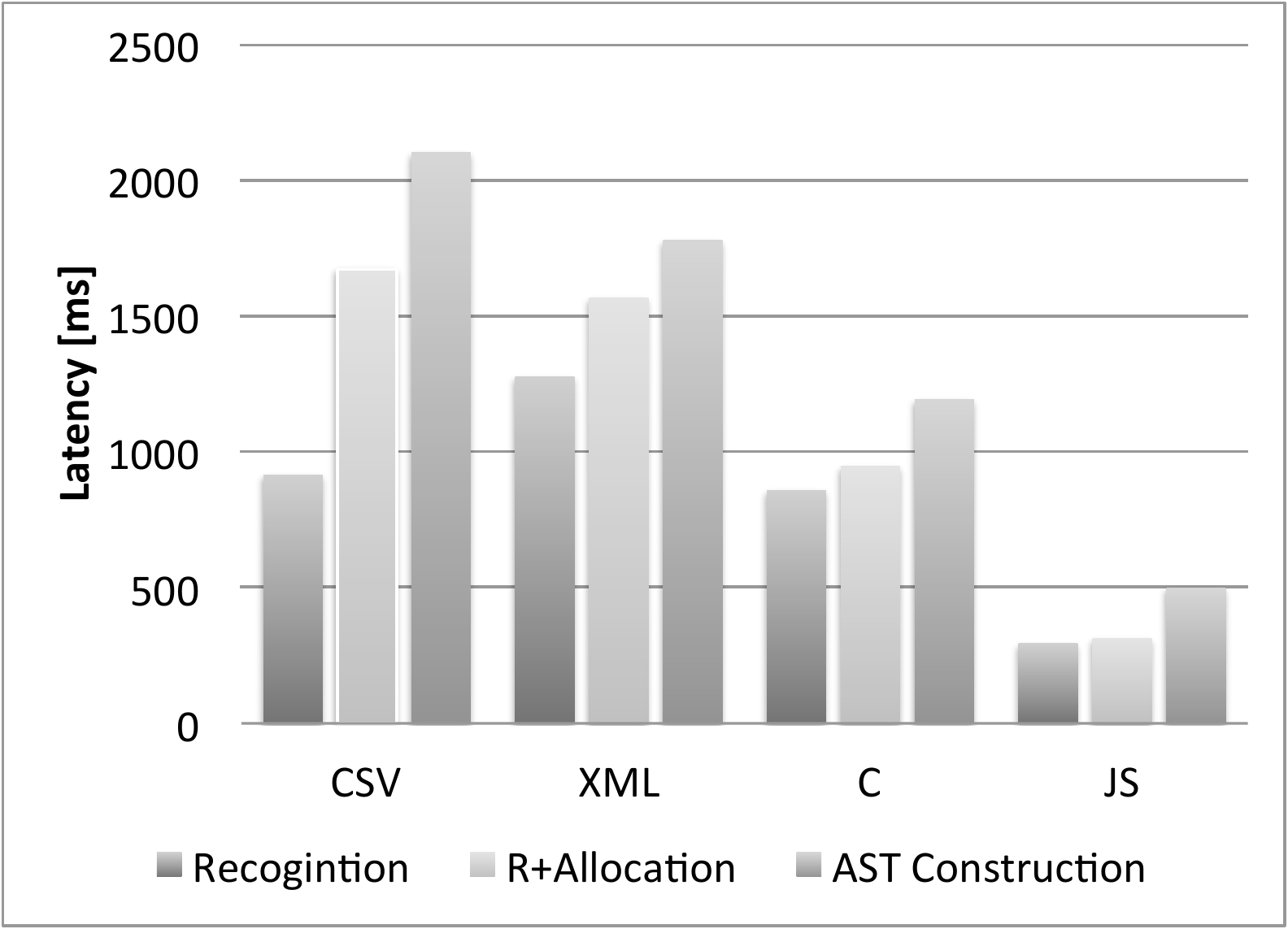}
\caption{Latency of AST Constructions in CSV, XML, C, JS}
\label{fig:latency}
\end{figure}

Now we will turn to the performance study. 
Figure \ref{fig:latency} shows the parsing time in each dataset. The data point labeled "Recognition" stands for paring time without AST construction, and "R+Allocation" stands for a cumulative time of "Recognition" and a simple instantiation time of AST nodes. The instantiation time is estimated by the elapsed time of the duplication of ASTs, whose size are the same as constructed in "AST Construction". It takes roughly 3 milliseconds to instantiate every 10,000 nodes. The differential time between "R+Allocation" and "AST Construction" implies a pure overhead of the transactional AST machine. We confirm that the transactional AST machine raises the time costs by 26\%, 16\%, 25\% and 59\% to the "R+Allocation" time in CSV, XML, C, and JS. The reason why the JS dataset shows the larger time cost may be the minor degradation of packrat parsing, which is indicated by the increased backtracking activity in Table \ref{table:grammars}.  

Table \ref {table:pegs} shows a performance comparison of other PEG-based parser generators. We have chosen Rats$!$ and PEG.js since they notably produce notably efficient parsers and are accepted in several third-party projects. Rats$!$ runs on Java8 as well as Nez, while PEG.js tested in the node.js environment including a V8-based JIT-compiler. To highlight the time cost of the underlying AST construction, we show the time difference between the "AST Construction" time and the "Recognition" time in millisecond. The experiment indicates that Rats$!$ is weak at parsing CSV and XML that contains many AST nodes. PEG.js shows good performance in total but is weak at parsing JavaScript that involves many backtracking. While the strength/weakness characteristic varies in datasets, Nez indicates the lowest time costs in all datasets. We confirm that the transactional AST machine achieves fast AST construction in contexts of PEG parsing. 

\begin{table}
\caption{AST construction time costs in millisecond  }
\label{table:pegs}
\begin{center}
\begin{tabular}{l|rrr} \hline
Grammar & Nez & Rats$!$ & PEG.js \\ \hline
CSV & 1,188 & 13,452 & 2,121 \\ 
XML & 510 & 7,766 & 1,466 \\
C & 205 & 390 & 469 \\
JS & 336 & 1,895 & 3,048 \\ \hline
\end{tabular}
\end{center}
\end{table}

\section{Related Work}


In a rich history of parser generators, many researchers have extended the construction of ASTs without semantic actions\cite{CACM92_Eli,LDTA10_ASTConstruction}. In total, our declarative approach has been inspired by SDF2 and Stratego/XT\cite{SCP08_Stratego,OOPSLA10_ParadiseLost}. ANTLR\cite{PLDI11_Antlr} provides both semantic actions and an additional support for AST construction, based on filtering from parse trees. These previous studies are not based on PEGs, but they suggest a substantial demand for declarative AST constructions in parser generators. 

Since Ford presented a formalism of PEGs\cite{POPL04_PEG}, many researchers and practitioners have been developed PEG-based parser generators: Leg/Peg (for C), Rats!\cite{PLDI06_Rats}, Mouse\cite{FI07_Mouse} (for Java), PEG.js (for JavaScript), and LPeg \cite{LPeg} (for Lua). Basically, these tools rely on language-dependent semantic actions for AST construction. Notably, LPeg provides the substring capturing, similarly to our approach, but other AST constructions can depend on semantic actions written in Lua programming languages. In semantic actions, the consistency management is the user's responsibility.

Waxeye\cite{Waxeye} is a unique exception in terms of unsupported semantic actions; it  provides automated AST construction based on filtering parse trees. Likewise, Rats! and some other PEGs tools provide similar options that enable filter-based tree constructions. However, the filtering parse tree is limited to the construction of the left-associative structure. 

\section{Conclusion}

This paper presented a declarative extension of PEGs for flexible AST constructions in such a way that AST can be transformed into nested trees, flattened lists, and left/right-associative pairs. The transactional AST machine is modeled to allow for the consistent AST construction with backtracking. In addition, the synchronous memoization is presented, integrating the packrat parsing to avoid potential exponential time costs. A transactional AST machine with the synchronous memoization is implemented in the Nez parser written in Java. We have demonstrated that the Nez parser requires a 25\% higher time cost for AST construction in most cases. In future work, we will investigate a more complex tree transformation with macro expansions while parsing.
   
\begin{acknowledgment}
The author thanks the attendees of JSSST Workshop on Programming and Programming Language 2015 for their feedback and discussions on our earlier work. 
\end{acknowledgment}

\bibliographystyle{ipsjsort-e}
\bibliography{../bib/parser,../bib/mypaper,../bib/url}  

\begin{biography}

\profile{Kimio Kuramitsu}{is an Associate Professor, leading the Software Assurance research group at Yokohama National University. His research interests range from programming language design, software engineering to data engineering, ubiquitous and dependable computing. He has received the Yamashita Memorial Research Award at IPSJ. His pedagogical achievements include Konoha and Aspen, the first programming exercise environment for novice students. He earned  his B. E. at the University of Tokyo, and his Ph. D. at the University of Tokyo under the supervision of Prof. Ken Sakamura.}

\end{biography}

\end{document}